\documentclass[twocolumn, superscriptaddress, nofootinbib, amsmath, amssymb,aps, floatfix, preprintnumbers,prd]{revtex4-2}
\usepackage[dvipsnames,table]{xcolor}
\usepackage{graphicx,epsfig,psfrag,bm,amssymb}
\usepackage[colorlinks=true,urlcolor=blue
,anchorcolor=blue,citecolor=blue,filecolor=blue,linkcolor=blue,menucolor=blue
]{hyperref}
\usepackage{dcolumn}
\usepackage{bm}
\usepackage{color}
\usepackage{soul}
\usepackage{cancel}
\usepackage{mathrsfs,amsfonts,color}
\usepackage{comment}
\usepackage{xspace}
\graphicspath{{figures/}}
\usepackage{cancel}
\usepackage{bbold}
\usepackage[export]{adjustbox}
\usepackage{siunitx}
\newcommand{\orcid}[1]{\begingroup
  \hypersetup{hidelinks}\href{https://orcid.org/#1}{\includegraphics[width=10pt]{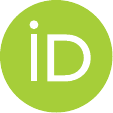}} \endgroup}

\newcommand{\TeV}{\mathrm{TeV}}
\newcommand{\GeV}{\mathrm{GeV}}
\newcommand{\MeV}{\mathrm{MeV}}
\newcommand{\eV}{\mathrm{eV}}
\newcommand{\keV}{\mathrm{keV}}
\newcommand{\m}{\mathrm{m}}
\newcommand{\ns}{\mathrm{ns}}

\newcommand{\dd}{d} 
\newcommand{\beq}{\begin{equation}}
\newcommand{\eeq}{\end{equation}}
\newcommand{\br}{\mathrm{BR}}
\newcommand{\hc}{\mathrm{h.c.}}
\newcommand{\cgg}{c_{GG}/f}
\newcommand{\fcgg}{\frac{c_{GG}}{f}}
\newcommand{\phion}{\ensuremath{\varphi\text{ion}}}
\newcommand{\phions}{\ensuremath{\varphi\text{ions}}}
\newcommand{\piH}{\ensuremath{\pi {\rm H}}}
\newcommand{\mgamc}{\textsc{MadGraph5\_aMC@NLO}}
\newcommand{\eg}{{e.\,g.}~}
\newcommand{\ie}{{i.\,e.}~}

\DeclareMathOperator\tr{tr}

\begin{document}

\title{Plugging the Pion Chimney}

\author{Nikita Blinov\orcid{0000-0002-2845-961X}}
\email{nblinov@yorku.ca}
\affiliation{Department of Physics and Astronomy, York University, Toronto, Ontario, M3J 1P3, Canada}

\author{Andrew M. Evans\orcid{0000-0002-6733-5599}}
\email{evansa3@uci.edu}
\affiliation{Department of Physics and Astronomy, University of California, Irvine, CA 92697}

\author{Shirley Weishi Li\orcid{0000-0002-2157-8982}}
\email{shirley.li@uci.edu}
\affiliation{Department of Physics and Astronomy, University of California, Irvine, CA 92697}

\date{November 16, 2026}

\begin{abstract}
Heavy QCD axions and other gluon-coupled axion-like particles (ALPs) with masses from MeV to GeV are well motivated and have been searched for extensively. However, a window around the neutral pion mass, the so-called pion chimney, has escaped every search. There the ALP mixes strongly with the $\pi^0$ and therefore shares its experimental signatures. We show that the chimney can be closed with existing LHC data. The key point is that mixing changes the lifetimes of the two mass eigenstates but does not make them equal, and at LHC energies the states are boosted enough for the lifetime to decide whether they escape the detector or decay inside it. The ATLAS monojet search constrains the first case and the ATLAS inclusive photon measurement the second, and together they exclude the chimney up to the cutoff of the ALP effective theory. Above the cutoff, LHC searches for the colored states of the simplest ultraviolet completions exclude the rest. We also show that the bound from the $K^+$ width does not reach the chimney, because an ALP passes the same selection as a $\pi^0$, and we derive an independent, purely kinematic constraint from the measured $\pi^\pm$--$\pi^0$ mass splitting.
\end{abstract}

\preprint{UCI-HEP-TR-2026-21}

\maketitle


\section{Introduction} 
 
Axions and axion-like particles (ALPs) are among the best motivated extensions of the Standard Model (SM).\footnote{We use ``axion'' and ``ALP'' interchangeably throughout this paper, even though we consider ALP mass and coupling to be independent parameters (see Sec.~\ref{sec:theory}).} The QCD axion solves the strong CP problem~\cite{Peccei:1977hh,Weinberg:1977ma,Wilczek:1977pj}, and its defining interaction is the coupling to gluons. The original axion is light and feebly coupled, but heavy variants, with masses from MeV to GeV and decay constants at the TeV scale, arise in models that address the axion quality problem~\cite{Kamionkowski:1992mf,Holman:1992us,Rubakov:1997vp,Berezhiani:2000gh,Hook:2014cda,Fukuda:2015ana,Gherghetta:2016fhp,Dimopoulos:2016lvn}. Such an ALP decays visibly, often promptly, and its gluon coupling makes it accessible in meson decays, beam dumps and hadron colliders~\cite{Beacham:2019nyx,Ilten:2022lfq}.

\begin{figure}
    \centering
    \includegraphics[width=0.45\textwidth]{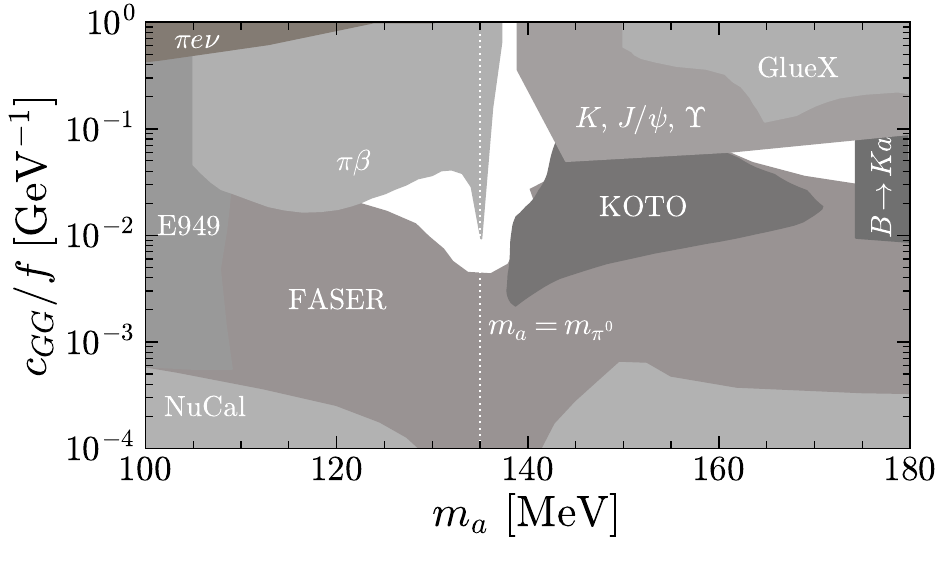}
    \caption{Existing constraints on a gluon-coupled ALP near the $\pi^0$ mass. Around $m_a = m_{\pi^0}$ (white dotted line), $a$--$\pi^0$ mixing makes the ALP decay promptly, and the existing bounds leave an unconstrained region, the pion chimney. Shown are FASER~\cite{FASER:2024bbl}, NuCal~\cite{Blumlein:1990ay,Dobrich:2019dxc,Jerhot:2022chi}, E949~\cite{E949:2005qiy,BNL-E949:2009dza}, KOTO~\cite{KOTO:2018dsc,Blinov:2021say}, NA62 (visible)~\cite{Blinov:2021say}, GlueX~\cite{Aloni:2019ruo}, $\pi\beta$ and $\pi e\nu$~\cite{Altmannshofer:2019yji}, $B \to K a$~\cite{Aloni:2018vki,Bertholet:2021hjl}, and $B \to K \mu^+\mu^-$ and $K, J/\psi, \Upsilon$~\cite{Bauer:2021mvw}.}
    \label{fig:pi-chimney}
\end{figure}

Over most of this mass range the constraints are extensive, but they fail near the neutral pion mass. A gluon-coupled ALP mixes with the $\pi^0$, and for an ALP $a$ with mass $m_a$ close to $m_{\pi^0}$ the mixing is strong even if the coupling is small. The ALP inherits the rapid two-photon decay of the $\pi^0$, while the state identified experimentally as the $\pi^0$ is itself partly ALP. As a result, the ALP is too short-lived for long-lived particle or missing-energy searches. Searches for visible decays are weakened as well. For $m_a < 3m_{\pi^0}$ the ALP decays dominantly to two photons, exactly like the $\pi^0$, and such searches typically exclude a mass window around $m_{\pi^0}$ from their analysis. What remains is an open region at $m_a \approx 120$--$150~\MeV$ and gluon coupling $\cgg \gtrsim 3\times10^{-3}~\GeV^{-1}$, shown in Fig.~\ref{fig:pi-chimney}, which has been called the pion chimney~\cite{Balkin:2025aer}.

The chimney has been noticed before. Reference~\cite{Goudzovski:2022vbt} singled out the window $100$--$200~\MeV$ as particularly difficult for kaon experiments, noted that the strong mixing there might be in tension with the measured $\pi^0$ lifetime, and called for a theoretical study of it. The same work derived the bound from the $K^+$ width that we revisit below. Reference~\cite{Balkin:2025aer}, which gave the chimney its name, proposed a recast of KOTO $K_L \to 3\pi^0$ calibration data and projected that future KOTO datasets could cover most of it.

In this paper we close the chimney. An ALP and a $\pi^0$ that mix strongly still differ in their lifetimes, and a lifetime can be measured once the particle is boosted. We therefore work with the exact mass eigenstates of the ALP--$\pi^0$ system, which remain well defined at $m_a = m_{\pi^0}$ where the perturbative mixing angle diverges, and turn to the Large Hadron Collider (LHC), where ALPs are produced with energies of hundreds of GeV. There the two eigenstates leave different signatures. One escapes the detector and appears as missing momentum recoiling against a jet, the other decays inside it and its collimated photons register as a single photon. The ATLAS monojet search~\cite{ATLAS:2021kxv} and the ATLAS inclusive photon measurement~\cite{ATLAS:2023yrt} constrain the two, and together they cover the chimney up to the cutoff of the ALP effective theory. Above the cutoff, the large couplings of the chimney imply new colored states light enough for LHC searches to exclude. We also reconsider the $K^+$ width bound, which does not survive a more careful analysis of the measurement which shows that an ALP passes the same experimental selection as a $\pi^0$. Finally, we derive an independent check of the narrowest part of the chimney from the measured $\pi^\pm$--$\pi^0$ mass splitting~\cite{Crawford:1990jc}, which depends only on two-body kinematics. Our results are summarized in Fig.~\ref{fig:ksvz-bound}.

Throughout we consider an ALP that couples only to gluons at the ultraviolet (UV) scale and work at leading order in the chiral expansion.

The paper is organized as follows. In Sec.~\ref{sec:theory}, we discuss our theoretical framework and $\pi^0-a$ mass diagonalization.  In Sec.~\ref{sec:kaon}, we show that the constraints from the $K^+$ decay width measurement are weakened when accounting for the detector acceptance for ALPs. In Sec.~\ref{sec:monojet}, we recast existing monojet searches, in Sec.~\ref{sec:inclusive-photon} we derive bounds from inclusive photon measurements, and in Sec.~\ref{sec:pi-mass} we describe constraints from the $\pi^0$ mass-splitting measurement. Lastly, in Sec.~\ref{sec:UV} we study the bounds from UV completions. We conclude in Sec.~\ref{sec:conclusion}.


\section{Theoretical Framework}
\label{sec:theory}

In this section, we set up the formalism we use to describe a gluon-coupled ALP with a mass near the neutral pion mass. We derive three quantities used in the rest of the paper, the mixing between ALP and $\pi^0$, the couplings of the resulting mass eigenstates to photons, and their couplings to nucleons. Our conventions follow Ref.~\cite{Blinov:2021say}, which builds on the chiral-rotation formalism of Refs.~\cite{Bauer:2020jbp,Bauer:2021wjo}.

We consider an ALP $a$ that couples to the Standard Model only through gluons,
\beq
 \mathcal{L} \supset \frac{1}{2}\left(\partial a\right)^2 - \frac{1}{2}m_a^2 a^2 + c_{GG} \frac{\alpha_s}{4 \pi} \frac{a}{f} G_{\mu \nu}^a \tilde{G}^{a, \mu \nu}
 \label{eq:fundamental_lagrangian}
\eeq
where $m_a$ is the ALP mass, $f$ is the ALP decay constant, and $c_{GG}$ is a dimensionless coupling. Low-energy observables depend on $c_{GG}$ and $f$ only through the combination $\cgg$, which is the coupling we constrain throughout. The decay constant appears by itself only as the cutoff of the effective theory, $\Lambda = 4\pi f$, when we discuss the validity of the EFT at large couplings (Secs.~\ref{sec:monojet} and \ref{sec:UV}). For a QCD axion the mass arises entirely from the gluon coupling, $m_a \approx m_\pi f_\pi\, (\cgg)$, which is roughly $0.1~\MeV$ for $\cgg = 10^{-2}~\GeV^{-1}$, well below the neutral pion mass. We therefore consider an ALP, whose mass term has an explicit contribution unrelated to QCD, and treat $m_a$ as a free parameter. The QCD contribution to the mass is still present, and we absorb it into the definition of $m_a$. Note that we do not include an explicit ALP-photon coupling in the UV.

At energies below the GeV scale, we cannot use Eq.~\eqref{eq:fundamental_lagrangian} directly since the relevant degrees of freedom are mesons and baryons rather than quarks and gluons. We therefore match this theory onto the chiral Lagrangian. The standard way to do this is to first remove the gluon operator by a chiral rotation of the quark fields,
\beq
 q \to e^{-i \kappa \gamma_5 \fcgg a} q,
\eeq
where $\kappa$ is a matrix in flavor space, diagonal in the quark-mass basis. This is an axial rotation of each quark field by an ALP-dependent phase, with opposite signs for its left- and right-handed components, and because of the chiral anomaly it is not a symmetry of the theory but shifts the $G\tilde{G}$ term. We work with two quark flavors, $q = (u, d)$, so $\kappa$ is a $2\times2$ matrix. This chiral rotation yields
\beq
\label{eq:rot-lag}
\begin{split}
 \mathcal{L} \supset &\left(1 - \tr \kappa \right) c_{GG} \frac{\alpha_s}{4 \pi} \frac{a}{f} G_{\mu \nu}^a \tilde{G}^{a, \mu \nu} + c_{\gamma \gamma} \frac{\alpha}{4 \pi} \frac{a}{f} F_{\mu \nu} \tilde{F}^{\mu \nu}\\
 &+c_{GG}\frac{\partial_\mu a}{f} \overline{q} \kappa \gamma^\mu \gamma_5 q - \overline{q} e^{-i \kappa \gamma_5 \fcgg a} m_q e^{-i \kappa \gamma_5 \fcgg a} q, 
\end{split}
\eeq
where $m_q$ is the quark mass matrix. The rotation trades the gluon coupling for three other terms: a coupling to photons, $c_{\gamma\gamma}$, generated by the chiral anomaly and given below, a derivative coupling to the quark axial currents, and an ALP-dependent quark mass matrix, $m_q(a) = e^{-i \kappa \gamma_5 \fcgg a} m_q e^{-i \kappa \gamma_5 \fcgg a}$. Choosing $\tr \kappa = 1$ removes the gluon term entirely. Besides the photon coupling, which carries over unchanged, the ALP then enters the chiral Lagrangian in two ways: through $m_q(a)$ in the mass term, and through the derivative coupling, which is matched onto the axial currents of the mesons and nucleons~\cite{Blinov:2021say}. The first is convenient because the chiral Lagrangian is already written in terms of the quark mass matrix, which is what gives the pions their mass, so the ALP couplings to pions follow by replacing $m_q$ with $m_q(a)$ in the known expressions, as we do below. The second gives the kinetic mixing and the nucleon couplings discussed below. Physical results cannot depend on $\kappa$ beyond the condition $\tr \kappa = 1$~\cite{Bauer:2020jbp,Blinov:2021say}. We take $\kappa = \frac{1}{2} \mathbb{1}$, which treats both flavors equally; the reason for this choice becomes clear in the next subsection. The coupling to photons is
\beq
 \label{eq:UV-photon-coupling}
 c_{\gamma \gamma} = -2 N_c c_{GG} \tr \left(Q^2 \kappa\right),
\eeq
where $Q$ is the quark charge matrix, $Q = \mathrm{diag}(2/3, -1/3)$, and $N_c = 3$ is the number of colors. For two flavors and $\kappa = \frac{1}{2} \mathbb{1}$, this gives $c_{\gamma\gamma} = -\frac{5}{3} c_{GG}$.


\subsection{Axion-Meson Mixing}

We now derive the mixing between the ALP and the neutral pion. In two-flavor chiral perturbation theory the pions enter through the matrix $\Sigma = \exp(2i\Pi/f_\pi)$, with $\Pi = \frac{1}{2}\begin{pmatrix}\pi^0 & \sqrt{2}\pi^+ \\ \sqrt{2}\pi^- & -\pi^0\end{pmatrix}$, and the quark masses enter through the leading-order mass term~\cite{Blinov:2021say}
\beq
\label{eq:chpt-mass-term}
 \mathcal{L} \supset \frac{f_\pi^2 B}{2} \tr\left[\Sigma\, m_q(a)^\dagger + m_q(a)\, \Sigma^\dagger\right],
\eeq
where $B = m_\pi^2/(m_u + m_d)$ is proportional to the quark condensate. The only change relative to the Standard Model is that $m_q$ is replaced by the ALP-dependent $m_q(a)$ of Eq.~\eqref{eq:rot-lag}. Inserting $m_q(a)$ into Eq.~\eqref{eq:chpt-mass-term} and expanding to quadratic order in the fields gives~\cite{Blinov:2021say}
\beq
 \mathcal{L} \supset-\frac{1}{2}m_{\pi}^{2}\left(\pi^{0}\right)^{2}-\frac{1}{2}m_{a}^{2} a^{2}+ \fcgg \delta m_{\pi}^{2} f_{\pi} a \pi^{0},
\eeq
where $f_\pi = 93~\MeV$ and $\delta=(m_{d}-m_{u})/(m_{d}+m_{u}) \approx 1/3$. The mixing term is proportional to $m_d - m_u$ for the following reason. With $\kappa \propto \mathbb{1}$, the ALP enters $m_q(a)$ as a flavor-singlet phase, so it can only mix with an isosinglet, whereas the $\pi^0$ is an isovector. Only the isospin breaking of the quark masses connects the two. The same expansion also produces the QCD contribution to the ALP mass mentioned above, $m_\pi^2 (c_{GG} f_\pi/f)^2$, which we have absorbed into $m_a$. For general $\kappa$ there is in addition a kinetic mixing term, proportional to $(\kappa_u - \kappa_d)\, \partial_\mu a\, \partial^\mu \pi^0$~\cite{Blinov:2021say}. Our choice $\kappa = \frac{1}{2}\mathbb{1}$ removes it, since the derivative coupling is then a flavor singlet, which does not couple to pions in the two-flavor theory, so the mixing is pure mass mixing and the Lagrangian above is complete at quadratic order. It can be written compactly in terms of a $\pi^0$--$a$ doublet $\mathcal{P} \equiv (\pi^0, a)^{\rm T}$,
\beq
\mathcal{L} \supset
 -\frac{1}{2}\mathcal{P}^{\rm T}M^2 \mathcal{P}.
\eeq
Here $M^2$ is the squared mass matrix given by
\beq
M^2 = m_{\pi}^{2}\begin{pmatrix}1 & -\epsilon \\ -\epsilon & 1 + \Delta\end{pmatrix}
\eeq
with $\epsilon = (\cgg) \delta f_{\pi}$ and $\Delta = m_a^2/m_\pi^2 - 1$. Away from the pion mass, this system is usually treated to first order in $f_\pi/f$~\cite{Bauer:2020jbp,Blinov:2021say}. In that approach one works in the flavor basis of $\pi^0$ and $a$ and treats the mixing term perturbatively. However, the mixing angle and the induced ALP couplings carry a factor $1/(m_\pi^2 - m_a^2)$ and diverge as $m_a \to m_\pi$, so the perturbative treatment cannot be used inside the pion chimney. We instead diagonalize $M^2$ exactly and work in the mass basis. $M^2$ is diagonalized by a $2\times 2$ orthogonal matrix, $R$, which is parametrized by a single mixing angle $\theta$, given by
\beq
\label{eq:mixing_angle}
 \tan \left(2 \theta\right) = \frac{2 \epsilon}{\Delta}.
\eeq
For $\epsilon \ll |\Delta|$ this reduces to the perturbative mixing angle $\theta \simeq \epsilon/\Delta$. The exact angle instead approaches maximal mixing, $\theta \to \pi/4$, at $\Delta = 0$. The physical states are the mass eigenstates, which propagate and decay independently of each other, and all observables below are expressed in terms of them. We will refer to the mass eigenstates as ``\phions'' $\varphi_{1,2}$ with $\varphi_i = R_{ij}\, \mathcal{P}_j$, with $R_{\pi 1} = R_{a2} = \cos\theta$, $R_{a1} = -R_{\pi 2} = -\sin\theta$; their masses are
\begin{align}
 m^2_{\varphi_1} &= m_\pi^2\left(1 + \frac{\Delta}{2} - \frac{1}{2}\sqrt{\Delta^2 + 4 \epsilon^2}\right),\\
 m^2_{\varphi_2} &= m_\pi^2\left(1 + \frac{\Delta}{2} + \frac{1}{2}\sqrt{\Delta^2 + 4 \epsilon^2}\right).
\end{align}
Since $\epsilon > 0$ the \phion{} masses are never degenerate, with $m_{\varphi_1} < m_{\varphi_2}$.
The mass splitting $\Delta m_\varphi$ is minimized when $\Delta = 0$:
\beq
 \label{eq:mass-splitting}
 \Delta m_\varphi^{\mathrm{min}} = m_\pi(\sqrt{1 + \epsilon} - \sqrt{1 - \epsilon}).
\eeq
The bottom of the pion chimney in Fig.~\ref{fig:pi-chimney} corresponds to $\cgg = 3\times10^{-3}~\GeV^{-1}$ or $\epsilon \simeq 10^{-4}$; this translates to a difference in masses of $\sim12.6~\keV$ between the two mass eigenstates. Although the mass splitting is small, it remains much larger than the intrinsic width of the $\pi^0$, about $8~\eV$, allowing the two mass eigenstates to be treated as distinct propagating states. This also means that we do not need to include any additional mass regulators near the resonance regions such as an imaginary part $i m_\pi \Gamma_\pi$ (as is often done for perturbative mixing calculations with heavier mesons~\cite{Ovchynnikov:2025gpx}). Exact diagonalization in this degenerate regime has also been carried out in Refs.~\cite{Cornella:2023kjq,DiLuzio:2024jip}, and up to differences in conventions our mixing angle and mass eigenvalues agree with theirs.


\subsection{Axion Interactions}
\label{sec:axion-interaction}

Every observable in this paper involves the production of one of the mass eigenstates and, in most cases, its decay to two photons. In this subsection we collect the couplings needed for both. They are obtained by rotating the interaction-basis couplings of the $\pi^0$ and the ALP into the \phion{} basis.

We begin with the decays. Both the $\pi^0$ and a light ALP ($m_a < 3m_\pi$) decay to two photons, and the mass eigenstates inherit a mixture of the two couplings. In the interaction basis, the couplings to photons are
\beq
\label{eq:photon-int-basis}
\mathcal{L} \supset \left( c_\pi\, \pi^0 + c_a\, a \right) F_{\mu\nu} \tilde{F}^{\mu\nu},
\eeq
where the neutral-pion coupling is fixed by the QED anomaly, while the ALP coupling follows from Eqs.~\eqref{eq:rot-lag} and \eqref{eq:UV-photon-coupling}, $c_a = c_{\gamma\gamma}\, \alpha/(4\pi f)$, giving
\beq
c_{\pi} = \frac{\alpha}{4\pi f_\pi}, \qquad c_{a} = -\frac{5\alpha c_{GG}}{12 \pi f}.
\eeq
Rotating to the mass eigenbasis gives $\mathcal{L} \supset \left( c_{\varphi_1} \varphi_1 + c_{\varphi_2} \varphi_2 \right) F_{\mu\nu} \tilde{F}^{\mu\nu}$ with $c_{\varphi_i} = R_{ij}\, c_j$, \ie $c_{\varphi_1} = c_\pi \cos\theta - c_a \sin\theta $ and $c_{\varphi_2} = c_a \cos\theta + c_\pi \sin\theta $, and the decay rates are
\beq
\label{eq:phion_decay_rate}
 \Gamma_{\varphi_i} = \frac{c_{\varphi_i}^2}{4\pi} m_{\varphi_i}^3.
\eeq
As a check, for $\epsilon \ll |\Delta|$ the photon coupling of the axion-like eigenstate reduces to $-(\alpha/4\pi)(\cgg)\left[5/3 - \delta/\Delta\right]$, which for $m_a \to 0$ gives the familiar $-(5/3 + \delta) \simeq -2$ (the ``$E/N - 1.92$'' of the QCD axion with $E=0$~\cite{DiLuzio:2020wdo}), and which vanishes at $\Delta \simeq 3\delta/5$, \ie $m_a \simeq 150~\MeV$. This cancellation will be visible in the LHC constraints of Secs.~\ref{sec:monojet} and \ref{sec:inclusive-photon}.

We next turn to production. At the LHC, the ALP is produced through the gluon operator of Eq.~\eqref{eq:fundamental_lagrangian} at energies far above the pion mass. The gluon operator produces the interaction-basis state $a$, which is a superposition of the mass eigenstates, $a = R_{a1}\varphi_1 + R_{a2}\varphi_2 = -\varphi_1 \sin\theta + \varphi_2 \cos\theta$. Each \phion{} is therefore produced with a weight $R_{ai}^2$ relative to the cross section of an unmixed ALP, and the photon couplings above then determine whether it decays inside the detector. We use this in Secs.~\ref{sec:monojet} and \ref{sec:inclusive-photon}. 

Next, the pion mass-splitting measurement of Sec.~\ref{sec:pi-mass} uses the charge-exchange reaction $\pi^- p \to n \pi^0$, so we also need the relative probability of emitting a \phion{} rather than a $\pi^0$ there. We take the effective proton interactions with the neutral pion and axion from Ref.~\cite{Blinov:2021say}. These follow from the leading-order pion-nucleon chiral Lagrangian, in which the pion couples to the nucleon through the isovector axial current with strength $g_A$, and the ALP, after the chiral rotation of Eq.~\eqref{eq:rot-lag}, couples through the flavor-singlet axial current with strength $g_0$. At tree level, the relevant derivative couplings are
\beq
\mathcal{L} \supset \bar{p} \gamma^{\mu} \gamma_{5}
\left( g_\pi \partial_{\mu} \pi^{0} + g_a \partial_{\mu} a \right) p,
\eeq
where
\beq
g_\pi = \frac{g_A}{2 f_\pi}, \qquad
g_a = \frac{g_0 c_{GG}\tr\kappa}{2 f}.
\eeq
Here, $g_A = 1.28$ and $g_0 = 0.384$ are the isovector and flavor-singlet nucleon axial couplings, respectively~\cite{Alexandrou:2020okk, Alexandrou:2021wzv}. Rotating $a$ and $\pi^0$ to the mass eigenbasis gives
\beq
\mathcal{L}\supset \bar{p} \gamma^{\mu} \gamma_{5}
\partial_\mu\left(
 g_{\varphi_1}\varphi_1
+ g_{\varphi_2} \varphi_{2}
\right)p,
\eeq
where $g_{\varphi_i} = R_{ij}\, g_j$, \ie $g_{\varphi_1} = g_\pi \cos\theta - g_a \sin\theta$ and $g_{\varphi_2} = g_a \cos\theta + g_\pi \sin\theta$. The amplitude for emitting $\varphi_i$ in charge exchange, relative to that for a SM $\pi^0$, is therefore $g_{\varphi_i}/g_\pi$, and the corresponding relative rate is $|g_{\varphi_i}/g_\pi|^2$. Finally, ALPs are also produced in rare meson decays, with amplitudes that follow from the same chiral Lagrangian~\cite{Bauer:2020jbp}. The relevant case, $K^+ \to \pi^+ a$, is discussed in Sec.~\ref{sec:kaon}.

With the mixing angle, the photon couplings, and the production weights in hand, we now turn to the experimental constraints.


\section{Kaon Decays}
\label{sec:kaon}

\begin{figure}
 \centering
 \includegraphics[width=0.45\textwidth]{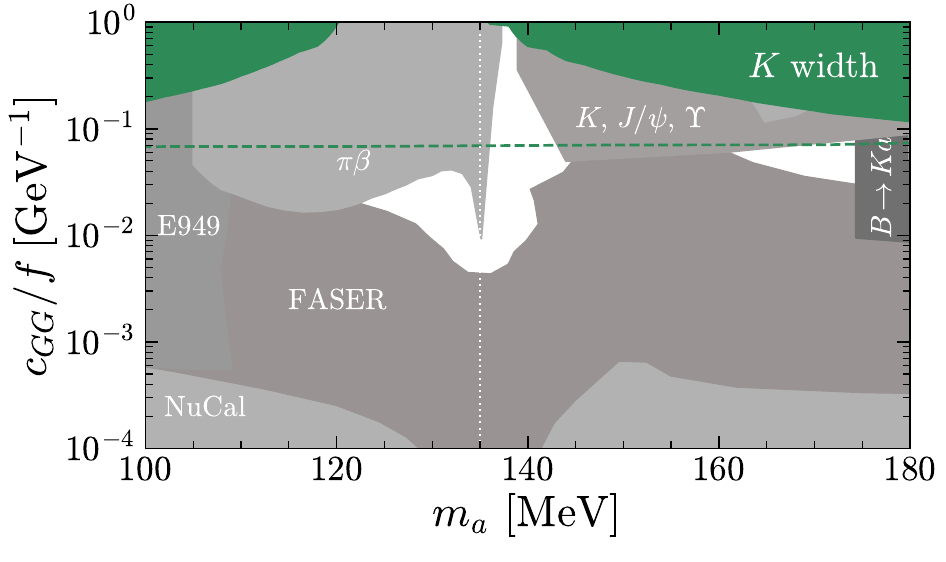}
 \caption{Gluon-coupled axion parameter space near the $\pi^0$ mass, with $m_a = m_{\pi^0}$ indicated by the white dotted line. Gray regions show existing limits summarized in Fig.~\ref{fig:pi-chimney}. The dashed line and the green region show the 95\%~CL constraint from $K^+$-width measurements~\cite{Goudzovski:2022vbt}, based on Ref.~\cite{KLOE:2008tnn}, before and after accounting for the finite mass resolution.}
 \label{fig:k-width-recast}
\end{figure}

In this section, we revisit the constraint from the measured $K^+$ branching fractions and total width, derived in Ref.~\cite{Goudzovski:2022vbt} and shown as the dashed line in Fig.~\ref{fig:k-width-recast}. Taken at face value it excludes the part of the pion chimney above $\cgg \sim 6\times10^{-2}~\GeV^{-1}$. The bound assumes that an ALP decay would show up as a missing piece of the kaon width. Near $m_a \approx m_{\pi^0}$, however, an ALP behaves so much like a neutral pion that this has to be taken into account. We show in this section that once it is, the bound weakens to the green region in Fig.~\ref{fig:k-width-recast} and no longer reaches the chimney.

The key argument of the bound is simple~\cite{Goudzovski:2022vbt}. The $K^+$ lifetime and the branching fractions of its exclusive decay modes are all measured, and the branching fractions sum to one within their uncertainties, $\br(K^+ \to \mathrm{SM}) = (100.00 \pm 0.15)\%$~\cite{ParticleDataGroup:2026aaa}. Any additional decay mode, such as $K^+ \to \pi^+ a$, has to fit inside that uncertainty, giving $\br(K^+ \to \pi^+ a) \lesssim 3 \times 10^{-3}$ at 95\% CL, and since $\Gamma(K^+ \to \pi^+ a) \propto (\cgg)^2$, this is the ceiling in Fig.~\ref{fig:k-width-recast}. The quoted uncertainty, however, comes from the PDG fit, which imposes that the Standard Model modes sum to exactly one. The proper test would be a fit without this constraint, which has not been done, so Ref.~\cite{Goudzovski:2022vbt} regards the bound as indicative. Our point is different. Even taking the number at face value, the bound does not apply near $m_{\pi^0}$.

We will argue that, in fact, these measurements do not provide a constraint in a relatively wide $\sim 100$ MeV window around $m_a \approx m_{\pi^0}$. The key point is that experimental measurements of branching fractions cannot distinguish between $a$ and $\pi$, \eg $K^+ \to \pi^+ \pi^0 (\gamma\gamma)$ and $K^+ \to \pi^+ a (\gamma\gamma)$, because the two are distinguished only through the $\gamma\gamma$ invariant mass, which has a finite resolution. This observation means that the reported measurement of $\br (\pi^+ \pi^0)^\mathrm{obs}$ must be interpreted as having two contributions:
\beq
\br (\pi^+ \pi^0)^\mathrm{obs} = \br (\pi^+ \pi^0) + \epsilon(m_a) \br (\pi^+ a)
\label{eq:kaon_br_obs}
\eeq
where $\epsilon(m_a)$ is the acceptance of the axion signal relative to $\pi^0$.

\begin{figure}
    \centering
    \includegraphics[width=0.45\textwidth]{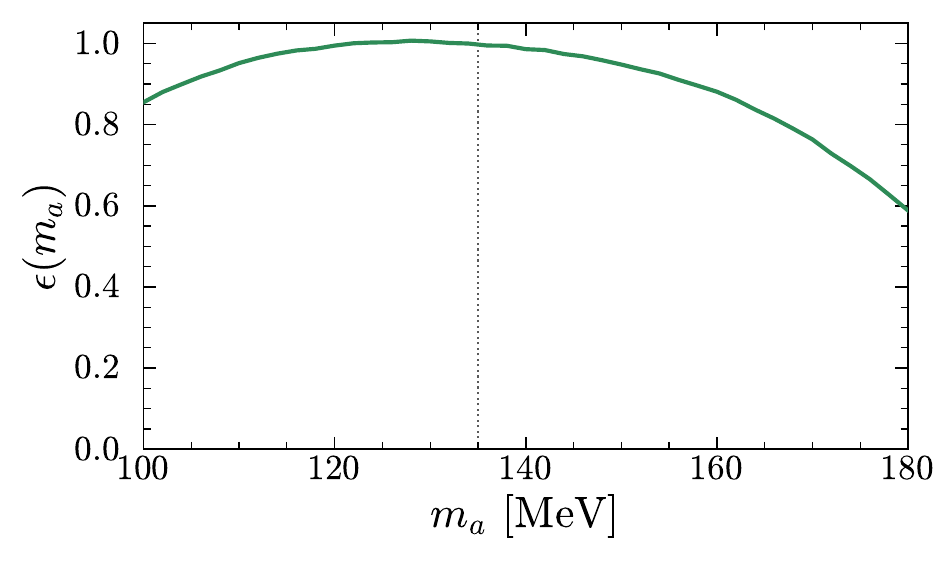}
    \caption{KLOE acceptance for $K^+ \to \pi^+ a$, $a \to \gamma\gamma$, relative to that for $K^+ \to \pi^+ \pi^0$, from our simulation of the KLOE selection with the $54~\MeV$ diphoton mass window. The dotted line marks $m_a = m_{\pi^0}$.}
    \label{fig:epsilon}
\end{figure}

The best measurement of $\br (K^+ \to \pi^+ \pi^0)^\mathrm{obs}$ comes from KLOE~\cite{KLOE:2008tnn}. There, the number of $K^+ \to \pi^+ \pi^0$ events is obtained from a fit to the $\pi^+$ momentum spectrum, in which the shape of the $K^+ \to \pi^+ \pi^0$ peak is taken from a control sample of events with a reconstructed $\pi^0$, selected by requiring two photons consistent in time and invariant mass with a $\pi^0$ from the kaon decay. An ALP that passes this selection enters the control sample together with the $\pi^0$ events, so the fit counts it as $K^+ \to \pi^+ \pi^0$. The acceptance $\epsilon(m_a)$ is therefore that of the control-sample selection for the ALP relative to the $\pi^0$. KLOE's default control sample applies no explicit cut on the diphoton mass. We adopt the $3\sigma \simeq 54\;\MeV$ mass window of their alternative control sample, which gives a conservative $\epsilon(m_a)$. We computed this by simulating KLOE events (production of $\phi$ nearly at rest, its decay to $K^+K^-$, followed by $K^+\to \pi^+ \pi^0(\gamma\gamma)$ or $\pi^+ a(\gamma\gamma)$) and comparing the acceptance for $\pi^0$ and $a$ using the selection criteria outlined in Ref.~\cite{KLOE:2008tnn}. The result is shown in Fig.~\ref{fig:epsilon}. As expected, when $m_{a} \approx m_{\pi^0}$, we find $\epsilon(m_a) \approx 1$. 

The axion decay channel modifies another observable~\cite{KLOE:2007wlh}, the total width of $K^+$:
\beq
\Gamma_{K^+}^\mathrm{obs} = \Gamma_{\text{other}} + \Gamma(\pi^+ \pi^0) + \Gamma(\pi^+ a), 
\label{eq:kaon_width}
\eeq
where $\Gamma_{\text{other}}$ is the sum of all other non-axion decay modes.\footnote{One can include the subdominant channels $K^+ \to \pi^+ \pi^0 \pi^0$ and $K^+ \to \pi^+ \pi^0 a$ into the analysis, but exactly the same considerations apply. Experimental selections on the final state $\pi^0$s are less strict compared to those in $K^+\to \pi^+ \pi^0$~\cite{KLOE:2003new}.}

Treating $\Gamma_{\text{other}}$ as known, Eqs.~\eqref{eq:kaon_br_obs} and \eqref{eq:kaon_width} are two linear equations for the two unknowns $\Gamma(\pi^+\pi^0)$ and $\Gamma(\pi^+ a)$, with solution
\beq
\Gamma(\pi^+a) = \frac{\left[(1-\br (\pi^+ \pi^0)^\mathrm{obs}) \Gamma_{K^+}^\mathrm{obs} - \Gamma_{\text{other}} \right]}{1- \epsilon(m_a)}. 
\eeq
At $\epsilon(m_a) = 1$ the two equations are degenerate, both depend only on $\Gamma(\pi^+\pi^0) + \Gamma(\pi^+ a)$, so an ALP contribution can always be absorbed into $\Gamma(\pi^+\pi^0)$ and there is no constraint at all. Only a theoretical prediction for $\Gamma(\pi^+\pi^0)$ could break this degeneracy, and hadronic uncertainties preclude one~\cite{Cirigliano:2011ny}. Away from $m_{\pi^0}$ the uncertainty on $\Gamma(\pi^+ a)$, and hence the bound on it, is inflated by $1/[1-\epsilon(m_a)]$. We therefore recast the ceiling of Ref.~\cite{Goudzovski:2022vbt} by scaling it with $1/\sqrt{1-\epsilon(m_a)}$, since $\Gamma(\pi^+ a) \propto (\cgg)^2$. 

Figure~\ref{fig:k-width-recast} compares the $K^+$ width measurement constraint before and after accounting for $\epsilon(m_a)$. The dashed line shows that without the $\epsilon(m_a)$ effect the width measurement eliminates all kinematically accessible parameter space above $\sim 6 \times 10^{-2}~\GeV^{-1}$. The solid green region shows that accounting for $\epsilon(m_a)$ weakens the bound throughout the plotted window and removes it entirely where $\epsilon(m_a) \approx 1$, roughly 120--135~MeV.

\section{ATLAS Monojet Search}
\label{sec:monojet}

We now turn to new constraints. The low-energy bound of the previous section fails near $m_{\pi^0}$ because an ALP there looks like a neutral pion. At the LHC the situation is different. The ALP is produced with an energy of hundreds of GeV, and the key question becomes whether each mass eigenstate decays inside the detector or escapes it. The two cases leave different signatures, missing transverse momentum or photons, and both have been searched for. In this section we consider the first.

The ATLAS collaboration has searched for ALPs in events with an energetic jet and large missing transverse momentum~\cite{ATLAS:2021kxv}.\footnote{The analogous CMS monojet analysis~\cite{CMS:2021far} does not provide an ALP interpretation of its results.} The idea is that an ALP produced in the hard collision escapes the detector, leaving missing transverse momentum, while the jet comes from the recoiling parton. The search excludes $\cgg > 10^{-3}~\GeV^{-1}$ for an ALP that escapes the detector, independent of its mass below $1~\GeV$.\footnote{ATLAS quotes the limit as $c_{\tilde G}/f_a < 8\times10^{-6}~\GeV^{-1}$ in the convention of Ref.~\cite{Brivio:2017ije}, which uses the Lagrangian $\mathcal{L} \supset -c_{\tilde G}\,(a/f_a)\, G\tilde{G}$ without the factor $\alpha_s/4\pi$, so $\cgg = (4\pi/\alpha_s)\, c_{\tilde G}/f_a \simeq 10^{-3}~\GeV^{-1}$ for $\alpha_s \simeq 0.1$.} 
If it applied, this would exclude the entire pion chimney, whose bottom lies at $3\times10^{-3}~\GeV^{-1}$.

We revisit this bound by examining when the ALP actually escapes the detector. The gluon operator produces the interaction-basis state $a = -\varphi_1 \sin\theta + \varphi_2 \cos\theta$ (Sec.~\ref{sec:axion-interaction}), and each mass eigenstate then travels with its own lifetime, set by its photon coupling. The ATLAS limit is a limit on the number of signal events. For an ALP that always escapes this number scales as $(\cgg)^2$, and the quoted coupling is the value at which it reaches the excluded rate. If only a fraction $P^{\rm inv}$ of the produced ALPs leave the detector before decaying, the signal rate is $(\cgg)^2 P^{\rm inv}$ instead, and the same excluded rate translates into the bound
\beq
    \fcgg \sqrt{ \sin^2\theta\, \langle P^{\rm inv}_1 \rangle + \cos^2\theta\, \langle P^{\rm inv}_2 \rangle} \leq 10^{-3}~\GeV^{-1}.
\label{eq:mj-bound}
\eeq
Here $\langle P^{\rm inv}_i\rangle$ is the survival probability of the $i$-th eigenstate averaged over the production kinematics. The two eigenstates are distinct particles, so their contributions add incoherently.

\begin{figure}
    \centering
    \includegraphics[width=0.45\textwidth]{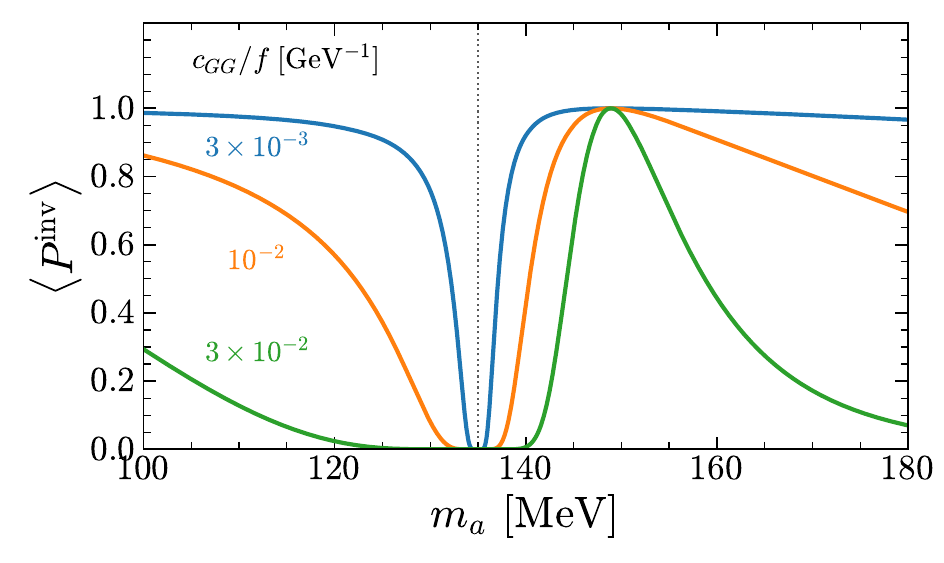}
    \caption{Probability that the state produced in $pp \to a + j$ escapes the ATLAS detector without decaying, averaged over the simulated events with $p_{T,j} > 400~\GeV$, as a function of $m_a$ for three values of $\cgg$. The dotted line marks $m_a = m_{\pi^0}$.}
    \label{fig:pinv}
\end{figure}

\begin{figure}[t]
    \centering
    \includegraphics[width=0.45\textwidth]{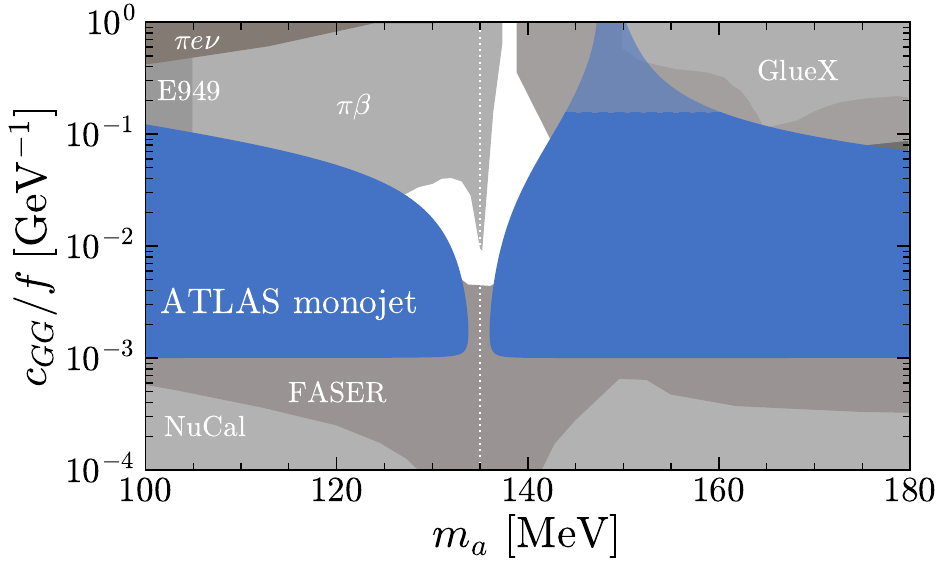}
    \caption{Gluon-coupled axion parameter space near the $\pi^0$ mass, with $m_a = m_{\pi^0}$ indicated by the white dotted line. Gray regions show existing limits summarized in Fig.~\ref{fig:pi-chimney}. The blue region shows our recast of the ATLAS monojet search~\cite{ATLAS:2021kxv}, requiring the produced state to travel $R_{\rm ATLAS} = 12.5~\m$ before decaying. The dashed line marks the EFT-validity limit $\cgg = 4\pi/\Lambda$ with $\Lambda = 1~\TeV$. Above it the region is drawn translucent, see text.}
    \label{fig:ATLAS-MJ}
\end{figure}

Whether an ALP escapes depends on how far it travels before decaying, which is set by its boost. We therefore need the energy spectrum of the produced ALPs, which we obtain by simulating $pp \to a + j$ at $\sqrt{s} = 13~\TeV$ at leading order using \mgamc~\cite{Alwall:2014hca}. We generate $10^5$ events with jet transverse momentum $p_{T,j} > 200~\GeV$ and jet pseudorapidity $|\eta_j| < 2.4$, matching the ATLAS signal region. At this order the ALP recoils against the jet, $p_{T,a} = p_{T,j}$. Since ATLAS reports that the sensitivity to the ALP signal is driven by the highest recoil bins, $p_T^{\rm recoil} > 400~\GeV$~\cite{ATLAS:2021kxv}, we evaluate the survival probability on the subsample of $N \simeq 10^4$ events with $p_{T,a} > 400~\GeV$, the part of the spectrum that sets the ATLAS limit. The averaged survival probability is then
\beq
    \langle P^{\rm inv}_i\rangle = \frac{1}{N} \sum_{k = 1}^{N} e^{-R_{\rm ATLAS}\Gamma_{i}/\gamma_{ik}} ,
\eeq
where $\Gamma_{i}$ is the decay rate of the $i$-th \phion{} given in Eq.~\eqref{eq:phion_decay_rate}, $R_{\rm ATLAS} = 12.5~\m$ is the characteristic radius required to escape the detector from the interaction point, which we take as a single effective distance independent of direction, and $\gamma_{ik} = E_k/m_{\varphi_i}$ is the Lorentz factor of $\varphi_i$ in the $k$-th simulated event, with $E_k$ the ALP energy in that event. 

Figure~\ref{fig:pinv} shows the quantity under the square root in Eq.~\eqref{eq:mj-bound}, which we denote $\langle P^{\rm inv}\rangle$, as a function of $m_a$ for three couplings. Even though it is a sum of two mass eigenstates, it is in practice dominated by one term: the pion-like eigenstate decays immediately and contributes nothing. For $m_a < m_{\pi^0}$ the lighter eigenstate $\varphi_1$ is the axion-like one and $\sin^2\theta \simeq 1$. Close to $m_{\pi^0}$ its pion admixture, $\theta \simeq \epsilon/\Delta$, is large enough that it decays inside the detector and $\langle P^{\rm inv}\rangle$ drops to zero. At fixed coupling $\langle P^{\rm inv}\rangle$ grows as $m_a$ moves away from the pion mass, because the admixture shrinks, and with it the pion-like contribution to the photon coupling. For $m_a > m_{\pi^0}$, $\varphi_2$ is the axion-like state and $\cos^2\theta \simeq 1$. Here the dependence on $m_a$ is not monotonic. The photon coupling of $\varphi_2$ is $c_a \cos\theta + c_\pi \sin\theta$, and since $\theta \simeq \epsilon/\Delta$ changes sign at $m_{\pi^0}$, the pion piece now subtracts from the ALP's own. The cancellation is complete at $150~\MeV$, as noted in Sec.~\ref{sec:axion-interaction}, where the state always escapes. Beyond it the ALP's own coupling wins again and $\langle P^{\rm inv}\rangle$ falls. 

Figure~\ref{fig:ATLAS-MJ} shows the recast ATLAS monojet constraint. Since Eq.~\eqref{eq:mj-bound} excludes a point when $\fcgg \sqrt{\langle P^{\rm inv}\rangle} > 10^{-3}~\GeV^{-1}$, the excluded region can be read off Fig.~\ref{fig:pinv}. Its lower boundary at $10^{-3}~\GeV^{-1}$ is the ATLAS limit for a state that escapes. The dip of $\langle P^{\rm inv}\rangle$ around $m_{\pi^0}$ is the chimney, and it widens with the coupling as the dip does. The upper boundary is where the ALP decays inside the detector, and the spike at $150~\MeV$, where it always escapes, is what remains excluded at the largest couplings.

There is a caveat to this constraint at large couplings: the validity of the EFT. The operator in Eq.~\eqref{eq:fundamental_lagrangian} is the leading term of an effective theory with cutoff $\Lambda = 4\pi f$.\footnote{This is the cutoff expected from naive dimensional analysis; a more precise version, the actual mass scale of new heavy particles, is considered in Sec.~\ref{sec:UV}.} At the LHC, the momentum flowing through the vertex is of order a TeV. If $f$ is at the TeV scale or below, this momentum is comparable to the cutoff and the operator no longer describes the physics, since the heavy states that generate it would be produced directly. We therefore mark $\cgg = 4\pi/\Lambda$ with $\Lambda = 1~\TeV$, i.e. $\cgg \simeq 0.15~\GeV^{-1}$, by a dashed line in Fig.~\ref{fig:ATLAS-MJ} and draw the excluded region above it translucent. Above this line we rely on the UV-completion arguments of Sec.~\ref{sec:UV}.


\section{ATLAS Inclusive Photon Measurements}
\label{sec:inclusive-photon}

We now turn to the second signature of Sec.~\ref{sec:monojet}, \phions{} that decay inside the detector. This is what happens in the gap around $m_{\pi^0}$ in Fig.~\ref{fig:ATLAS-MJ}, where the produced state decays before escaping. These decays are visible in the ATLAS inclusive photon cross section data~\cite{ATLAS:2023yrt}, because the \phions{} produced at the LHC are so boosted that the two decay photons merge and are reconstructed as a single photon.\footnote{CMS inclusive photon measurements use a smaller dataset at the moment~\cite{CMS:2018qao,CMS-PAS-SMP-25-004}.} Unlike the $\pi^0$s inside jets, a \phion{} recoiling against the jet is isolated, so it enters the isolated-photon measurement. As we will see, the measurement is sensitive to this signal because its spectrum is harder than the SM background.

To predict the photon signal we need the energy and direction of each produced \phion{} and the probability that it decays before reaching the calorimeter. We obtain the former using \mgamc~\cite{Alwall:2014hca}, simulating $N = 2\times10^5$ $pp \to a + j$ events at $\sqrt{s} = 13~\TeV$, generated with $p_{T,j} > 200~\GeV$, to determine a differential cross section in produced photon transverse energy and pseudorapidity, $\dd^2 \sigma/\dd E_T^\gamma \dd \eta^\gamma$, from axion decays. 

For a produced \phion{} to register as a photon, two things must happen. First, it must decay before reaching the ATLAS electromagnetic calorimeter (ECAL). Second, it must be boosted enough that its two photons appear as a single photon in the detector. The first requirement is fulfilled by weighting the simulated events by the average decay probability over the flight distance $L_k$ to the calorimeter front face, which depends on the direction of the \phion{} (the barrel at a radius of $1.5~\m$ or the endcap at $|z| = 3.7~\m$),
\beq
  \langle P^{\rm decay}_i \rangle = \frac{1}{N}\sum_{k = 1}^N \left(1 - e^{-L_k \Gamma_{i}/\gamma_{ik}}\right),
\eeq
where $L_k$, $\Gamma_i$ and $\gamma_{ik}$ are the flight distance, decay rate and Lorentz factor, as in Sec.~\ref{sec:monojet}. The produced state is again the superposition $a = -\varphi_1\sin\theta + \varphi_2\cos\theta$, so the signal is the simulated cross section times $\sin^2\theta\, \langle P^{\rm decay}_1\rangle + \cos^2\theta\, \langle P^{\rm decay}_2\rangle$. Due to the large $p_{T,a}$ and comparatively small $m_{\varphi_i}$ we find that the separation of photons produced in \phion{} decays, roughly $2 m_{\varphi_i}/E \lesssim 1~\mathrm{mrad}$ for $E_T > 250~\GeV$, is always below the ECAL angular resolution, whose finest segmentation is about $0.003$ in $\eta$. Thus the second condition is always satisfied across our simulated events.

ATLAS unfolds its measured photon spectrum for detector effects and reports it as a particle-level cross section in a specified fiducial region~\cite{ATLAS:2023yrt}. Our simulated rate can therefore be compared to the data directly, without modeling the detector, so long as the photons from the axion decay are recorded as a single photon, as argued above. We apply the same kinematic requirements as the measurement: transverse energy $E_T^\gamma > 250~\GeV$ and pseudorapidity $|\eta| < 2.37$, excluding the gap between the central and forward calorimeters at $1.37 < |\eta| < 1.56$. 

The result of our simulation is a predicted \phion{} contribution to the isolated-photon spectrum, binned in $E_T^\gamma$ and $|\eta^\gamma|$ exactly as in the measurement. We add the predicted \phion{} spectrum to the next-to-next-to-leading-order QCD prediction (NNLOJET) presented in~\cite{ATLAS:2023yrt} and perform a $\chi^2$ test against data. Our $\chi^2$ takes the form
\beq
  \chi^2 = R_m C^{-1}_{m n} R_n,
\eeq
where $R_n = D_n - (T_n + S_n(m_a, \cgg))$ with $D_n$ being the data, $T_n$ being the NNLOJET prediction, and $S_n(m_a, \cgg)$ being the \phion{} signal in the $n$-th bin.

ATLAS does not provide a covariance matrix but the size of each uncertainty component in every bin, together with its correlation between bins~\cite{ATLAS:2023yrt}. We build $C_{mn}$ from these. Statistical and several detector uncertainties are uncorrelated between bins, the photon energy-scale and resolution components are each fully correlated across bins, and the theory scale uncertainty, whose correlation ATLAS leaves unspecified, we take as correlated. We add the statistical uncertainty of our simulated signal, and we exclude bins in which the simulated signal has fewer than 50 events or the measurement uncertainty exceeds 25\%, where the Gaussian $\chi^2$ is not reliable. For each $m_a$ we define $\Delta\chi^2(\cgg) = \chi^2(m_a, \cgg) - \chi^2_{\rm min}(m_a)$, where $\chi^2_{\rm min}$ is the minimum over $\cgg \geq 0$, and set a one-sided 95\% CL upper limit at the coupling $(\cgg)_{95}$ where $\Delta\chi^2((\cgg)_{95}) = 2.706$.

\begin{figure}
  \centering
  \includegraphics[width=0.45\textwidth]{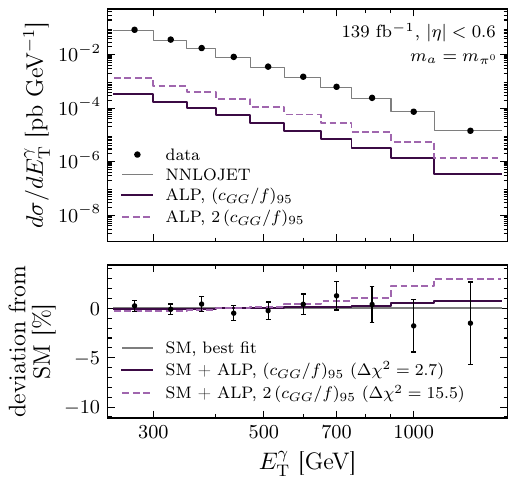}
  \caption{Isolated-photon transverse-energy spectrum for $|\eta|<0.6$, the most sensitive $|\eta|$ region of the ATLAS measurement at $\sqrt{s}=13$~TeV with 139~fb$^{-1}$~\cite{ATLAS:2023yrt}. Upper: measured cross section, the NNLOJET prediction, and the ALP contribution at $m_a = m_{\pi^0}$ for $c_{GG}/f = (c_{GG}/f)_{95}$, the 95\% CL upper limit, and $2\,(c_{GG}/f)_{95}$. Lower: relative difference, in percent, from the best-fit SM prediction, i.e. NNLOJET with the correlated experimental and theory uncertainties set to their best-fit values. Data error bars show the uncorrelated uncertainty only. $\Delta\chi^2$ values are for the combined fit to all $|\eta|$ regions.}
  \label{fig:ppja_spectrum}
\end{figure}

Figure~\ref{fig:ppja_spectrum} shows the measured spectrum in the most sensitive region, $|\eta|<0.6$, together with the NNLOJET prediction and the \phion{} contribution at the resulting limit. The \phion{} spectrum is harder than the SM one, so its relative contribution grows with $E_T^\gamma$. The lower panel shows data and predictions relative to the best-fit SM. The data scatter around zero, while the \phion{} adds a contribution that grows with $E_T^\gamma$ and cannot be absorbed by an overall shift in normalization, so the SM + \phion{} curves depart from the data at high $E_T^\gamma$.

\begin{figure}
  \centering
  \includegraphics[width=0.45\textwidth]{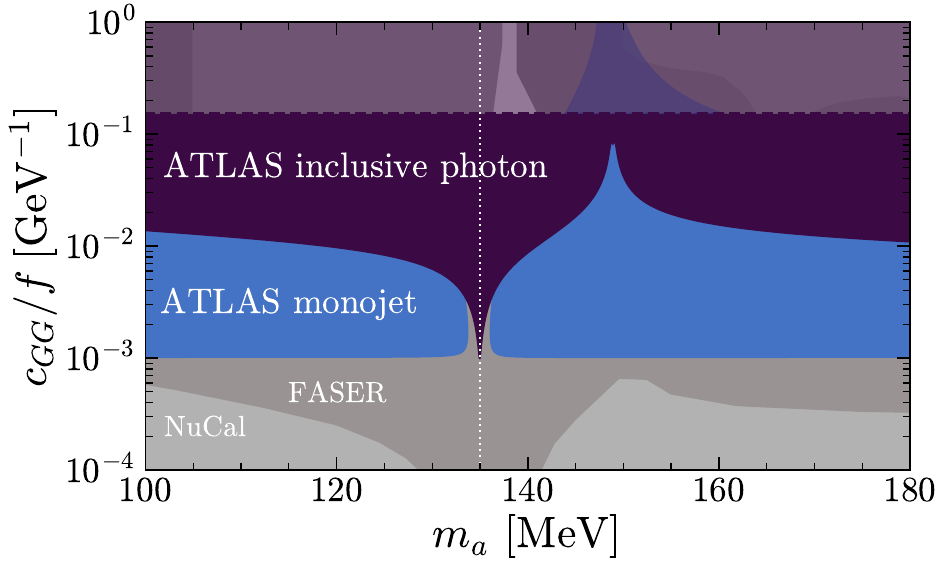}
  \caption{Gluon-coupled axion parameter space near the $\pi^0$ mass, with $m_a = m_{\pi^0}$ indicated by the white dotted line. Gray regions show existing limits summarized in Fig.~\ref{fig:pi-chimney}, while the blue region shows the ATLAS monojet constraint of Fig.~\ref{fig:ATLAS-MJ}. The purple region shows the 95\%~CL constraint from $pp \to a(\gamma\gamma) + j$, derived from the ATLAS inclusive isolated-photon measurement~\cite{ATLAS:2023yrt} (Sec.~\ref{sec:inclusive-photon}), where the decay $\varphi_i \to \gamma\gamma$ mimics a single photon. The dashed line marks the EFT-validity limit $\cgg = 4\pi/\Lambda$ with $\Lambda = 1~\TeV$. Above it both LHC regions are drawn translucent, see text.}
  \label{fig:ATLAS-inclusive}
\end{figure}

The constraint from the ATLAS inclusive photon measurement is presented in Fig.~\ref{fig:ATLAS-inclusive}. In contrast to the monojet constraint, this one is enhanced by $a$--$\pi^0$ mixing, which increases the probability for the long-lived state to decay within the detector and produce a visible photon signal. In particular, the region around $m_{\pi^0}$ that the monojet search cannot reach (Fig.~\ref{fig:ATLAS-MJ}) is excluded here, because there the produced \phion{} decays promptly and contributes photons. The two LHC searches are therefore complementary, and together they cover the chimney up to the EFT cutoff. The spike to the right of the chimney, at $m_a \simeq 150~\MeV$, arises due to an accidental cancellation in the photon coupling of the long-lived mass eigenstate, as discussed in Sec.~\ref{sec:monojet}. The lower boundary of the constraint is set by the predicted signal becoming too small to be distinguished from the background.

The EFT caveat of Sec.~\ref{sec:monojet} applies here as well. The purple region is drawn solid up to $\cgg = 4\pi/\Lambda$ with $\Lambda = 1~\TeV$, marked by the dashed line in Fig.~\ref{fig:ATLAS-inclusive}, and translucent above it.


\section{Pion Mass Splitting Measurement}
\label{sec:pi-mass}

The pion chimney can also be probed by the measurement of the $\pi^\pm$--$\pi^0$ mass splitting, which still provides the most precise determination of the $\pi^0$ mass~\cite{ParticleDataGroup:2026aaa} but is rarely used to probe physics beyond the SM. Reference~\cite{Crawford:1990jc} measured it from the speed of the neutron in the charge-exchange decay of pionic hydrogen at rest, $\piH \to n\pi^0$, whose two-body kinematics fix the neutron speed in terms of the masses, using the neutron time of flight at baselines of $3.2$, $7.9$ and $18.1~\m$. With $a$--$\pi^0$ mixing, \piH{} decays to either mass eigenstate, and each gives the neutron a slightly different speed, so the single neutron peak becomes two peaks separated in arrival time by about $17~\mathrm{ns}\,(L/1~\m)(\Delta m_\varphi/\MeV)$ (Appendix~\ref{app:tof}).

The published spectra cover only $\pm 30~\mathrm{ns}$ around the peak, and the overall normalization is not predicted, so a second peak outside the window simply goes unrecorded. The measurement is therefore sensitive only when both peaks fall inside the window, which requires $|m_a - m_{\pi^0}| \lesssim 0.6~\MeV$ at the shortest baseline and $\lesssim 0.1~\MeV$ at the longest, and only for $\cgg \lesssim 0.1~\GeV^{-1}$, above which the mixing itself pushes the two eigenstates too far apart.

We quantify the sensitivity with a binned likelihood fit of the two-peak spectrum to the published spectra at the three baselines, taking the fitted single-peak curve of Ref.~\cite{Crawford:1990jc} as the template and treating the arrival-time offset as a free parameter (see Appendix~\ref{app:tof} for details). The region excluded at 95\% CL, shown in Fig.~\ref{fig:ksvz-bound}, is $134.1~\MeV \lesssim m_a \lesssim 135.4~\MeV$ and $10^{-3}~\GeV^{-1} \lesssim \cgg \lesssim 0.1~\GeV^{-1}$. It lies entirely inside the region excluded by the inclusive photon measurement of Sec.~\ref{sec:inclusive-photon}, so it adds no constraints, but it is an independent check of the narrowest part of the chimney that relies on nothing but two-body kinematics.


\section{Constraints from UV Completions}
\label{sec:UV}

In addition to direct searches for ALP production, the pion chimney parameter space is amenable to indirect probes. Here we consider the production of the heavy physics responsible for the $a G\widetilde{G}$ operator at the LHC. This is less robust than the direct constraints of the previous sections, since it depends on the unknown UV physics, but it is needed nonetheless. The direct bounds of Secs.~\ref{sec:monojet} and \ref{sec:inclusive-photon} were capped at $\cgg \simeq 0.15~\GeV^{-1}$, where the axion EFT breaks down, and the arguments below constrain the region above the cap.

The scale $f$ that determines the axion-gluon coupling is related to the mass scale of UV physics that completes the dimension-5 operator $a G\widetilde{G}$. In particular, large $\cgg$ values in the pion chimney imply a low scale of the UV completion, low enough for the LHC to test it directly.  

The simplest UV completions of our minimal $aG\widetilde{G}$ model with a gluon-only interaction are Kim–Shifman–Vainshtein–Zakharov (KSVZ)-type axion models~\cite{DiLuzio:2020wdo}.\footnote{A DFSZ-type completion, with additional Higgs doublets and no new quarks, evades the bound altogether, but it generates ALP couplings to the SM quarks and lies outside the gluon-only model considered here.} In these scenarios the UV consists of fermions $Q$ in a vector-like representation of the SM gauge group, and a scalar $S$. A non-trivial representation under $SU(3)_C$ is mandatory to generate the $aG\widetilde{G}$ coupling, while charges under the electroweak group are optional. These fields also carry a global Peccei-Quinn (PQ) charge, which forbids a bare mass for $Q$ but allows the Yukawa interaction
\beq
  \mathcal{L}\supset - y_Q S Q_L Q_R^c + \hc
  \label{eq:ksvz_lang}
\eeq
Assuming the scalar potential $V(S)$ has a minimum at $\langle S \rangle = (v_S/\sqrt{2}) \exp(i a /v_S)$, PQ symmetry breaking generates a mass term for the heavy quarks
\beq
  \mathcal{L}\supset  - m_Q e^{i a /v_{S}} Q_L Q_R^c + \hc ,\; m_Q = \frac{y_Q v_S}{\sqrt{2}}.
  \label{eq:mass_term_after_pq}
\eeq
A chiral rotation eliminates the axion from the mass term and produces the axion-gluon coupling
\beq
  \Delta \mathcal{L} = \frac{\alpha_s N}{4\pi} \frac{a}{v_S}G\widetilde{G}
  \label{eq:axion_shift_to_ggdual}
\eeq
where $N$ is the color anomaly coefficient, set by the representation of $Q_{L,R}$ under the SM gauge group; for an $SU(3)_C$ fundamental with no other charges $N = 1/2$, the Dynkin index of the representation. Comparing Eq.~\eqref{eq:fundamental_lagrangian} and Eq.~\eqref{eq:axion_shift_to_ggdual} we see that\footnote{Another common convention is to define the axion scale $f$ via $f \equiv v_S/(2N)$, as in Ref.~\cite{DiLuzio:2020wdo}; we instead follow Ref.~\cite{Bauer:2021mvw}.}
\beq
  \fcgg = \frac{N}{v_S}.
\eeq
This relation, together with the heavy quark mass in Eq.~\eqref{eq:mass_term_after_pq} relates the mass of the new states to the axion coupling:
\beq
  m_Q = \frac{y_Q N}{\sqrt{2} \cgg}. 
  \label{eq:mass_in_terms_of_cgg}
\eeq
In Appendix~\ref{sec:UV-robust} we show that this relation cannot be evaded by a less minimal KSVZ sector.

The LHC strongly constrains new QCD-charged matter, so through Eq.~\eqref{eq:mass_in_terms_of_cgg} a lower bound on $m_Q$ becomes an upper bound on $\cgg$, once $N$ and $y_Q$ are bounded. Both are bounded by theoretical arguments. First, $N$ roughly counts how many additional SM-interacting states are added to the theory, since the Dynkin index grows with the size of the representation~\cite{DiLuzio:2017pfr}. These states modify the running of the SM gauge couplings and can lead to Landau poles. Second, if the reheating temperature is above their mass, the heavy quarks are produced in the early universe and freeze out with a substantial abundance. For certain charge assignments the resulting hadrons carry fractional charge and cannot decay to SM particles, so they survive to the present day,\footnote{In particular, this eliminates the most basic model with no hypercharge assignment.} and searches for fractionally charged particles strongly constrain them. This can be avoided if the heavy quarks have non-gauge interactions with light SM particles that let them decay. These considerations limit the possible choices of $N$. They were explored in Refs.~\cite{DiLuzio:2016sbl,DiLuzio:2017pfr} and summarized in Ref.~\cite{DiLuzio:2020wdo}; Ref.~\cite{DiLuzio:2016sbl} finds that only the window $1/2 \leq N \leq 10$ is viable. There are more arguments to prefer lower $N$. For $2N > 1$ the low-energy axion potential has multiple degenerate minima~\cite{GrillidiCortona:2015jxo}, which leads to the axion domain wall problem if those minima are populated in the early universe.
 
Finally we consider $y_Q$. A naive perturbativity constraint is $y_Q < 4\pi$, but a reasonable argument strengthens it considerably. Consider the simplest viable charge assignment from the arguments above, a heavy quark with the quantum numbers of a right-handed down quark, $Q_L \sim (3,1,-1/3)$ under $SU(3)_C\times SU(2)_L \times U(1)_Y$~\cite{DiLuzio:2016sbl}. The one-loop RGE for the Yukawa coupling in Eq.~\eqref{eq:ksvz_lang} is\footnote{We used  \href{https://github.com/LSartore/pyrate/tree/master}{\texttt{pyr@te}} to generate the RGEs~\cite{Sartore:2020gou}.}
\beq
\frac{dy_Q}{d\ln\mu} = \frac{y_Q}{16\pi^2}\left( 4y_Q^2 - 8g_3^2 - \dots \right)
\label{eq:ksvz_rge}
\eeq
where the dots stand for $U(1)_Y$ terms, which we neglect. The two terms compete. The gauge term, $-8g_3^2$, drives $y_Q$ down as $\mu$ grows, while the self-coupling term, $4y_Q^2$, drives it up. At the weak scale $\alpha_s \sim 0.1$, so $g_3 \sim 1.1$. Suppose we start with the largest perturbative value, $y_Q \equiv y_0 = 4\pi$, at the input scale $\mu_0$, the heavy quark mass. The self-coupling term then dominates by a factor of sixty, and dropping the gauge term the RGE solves to
\beq
y_Q(t) \approx \frac{y_0}{\sqrt{1-t y_0^2/(2\pi^2)}},
\eeq
with $t= \ln \mu/\mu_0$. The solution diverges at $t = 2\pi^2/y_0^2 = 1/8$, that is at $\mu \approx 1.1\,\mu_0$, so the UV theory is barely valid above its own mass scale. This is an extreme case, but the behavior is qualitatively the same as long as $4y_Q^2 - 8g_3^2 > 0$. The strong coupling term merely shifts the Landau pole to somewhat higher energies. For example, if we treat $g_3$ as a constant, the Landau pole moves to
\beq
    \mu = \mu_0 \left(\frac{y_0^2}{y_0^2 - 2g_3^2}\right)^{\pi^2/g_3^2}.
\eeq
If the UV theory is to have an extended range of validity, as one would want, then $y_0^2 \approx 2g_3^2$ must be fine-tuned, or, by Eq.~\eqref{eq:ksvz_rge}, $4y_Q^2 \lesssim 8g_3^2$, so that there is no Landau pole at all. Roughly this translates to
\beq
y_Q \lesssim \sqrt{8\pi \alpha_s} = 1.6 \sqrt{\frac{\alpha_s}{0.1}},
\eeq
with $\alpha_s$ evaluated at the weak scale. This is a much stronger condition than $y_Q < 4\pi$.

\begin{figure}
    \centering
    \includegraphics[width=0.45\textwidth]{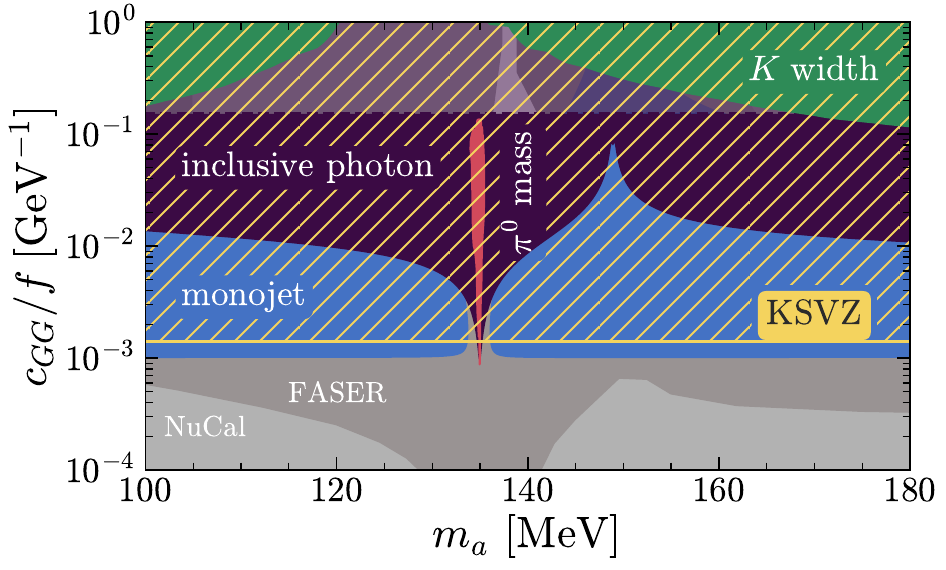}
    \caption{Gluon-coupled axion parameter space near the $\pi^0$ mass. Gray regions show existing limits summarized in Fig.~\ref{fig:pi-chimney}. The remaining regions are derived in this work: the ATLAS monojet search (blue, ``ATLAS monojet'', Sec.~\ref{sec:monojet}), the ATLAS inclusive isolated-photon measurement (purple, ``ATLAS inclusive photon'', Sec.~\ref{sec:inclusive-photon}), the $K^+$ width (green, Sec.~\ref{sec:kaon}), the neutron time-of-flight measurement of $m_{\pi^\pm} - m_{\pi^0}$ (pink, ``$\pi^0$ mass'', Sec.~\ref{sec:pi-mass}), and bounds on the KSVZ model (hashed yellow, Sec.~\ref{sec:UV}).}
    \label{fig:ksvz-bound}
\end{figure}

The LHC excludes new colored particles below about a TeV; see Ref.~\cite{Arhrib:2026coy} for a recent compilation of bounds on vector-like quarks and the review~\cite{Banerjee:2024zvg}. The bounds depend somewhat on the specific representations and decay modes, but they generally demand that $m_Q \gtrsim 1 $ TeV. Using Eq.~\eqref{eq:mass_in_terms_of_cgg}, we obtain a rough bound on $\cgg$:
\beq
\label{eq:ksvz}
  \fcgg < 1.4\times 10^{-3}\;\GeV^{-1} \; \left(\frac{N y_Q}{2}\right).
\eeq
For $N y_Q = 2$ Eq.~\eqref{eq:ksvz} lies at $1.4\times10^{-3}~\GeV^{-1}$, below the bottom of the chimney, so the simplest KSVZ completion is excluded throughout it. This is the constraint we draw in Fig.~\ref{fig:ksvz-bound}. The most generous choice allowed by the arguments above, $N = 10$ and $y_Q \simeq 1.6$, raises the bound by a factor of $8$, to $\cgg \simeq 10^{-2}~\GeV^{-1}$. Whatever the completion, the upper part of the chimney, including the region above the EFT cap of Secs.~\ref{sec:monojet} and \ref{sec:inclusive-photon}, requires colored states that the LHC has excluded.

We have not used another indirect probe, the chromomagnetic moment of the top quark generated by the renormalization group running of the $aG\widetilde{G}$ operator~\cite{Galda:2021hbr,Bauer:2021mvw}, which ATLAS and CMS constrain~\cite{ATLAS:2023eld,CMS:2025ugn}. For $\Lambda = 1~\TeV$ it would only reach $\cgg \gtrsim 0.5~\GeV^{-1}$, far inside the region where the EFT that the running relies on has already broken down.


\section{Conclusion}
\label{sec:conclusion}

Near $m_{\pi^0}$ a gluon-coupled ALP and the neutral pion form a strongly mixed system, and treating them as unrelated particles can lead to a misinterpretation of experimental results. We diagonalized the $a$--$\pi^0$ mass matrix exactly and formulated every observable in terms of the two mass eigenstates, so that nothing relies on the small-mixing angle approximation that fails at $m_a \approx m_{\pi^0}$. The observations that close the chimney are simple. The mixing changes the lifetimes of the two states but does not erase the difference between them, and at the LHC the states are boosted enough for that difference to show up, whether they decay inside the detector or escape it.

The results are summarized in Fig.~\ref{fig:ksvz-bound}. The $K^+$ width bound of Ref.~\cite{Goudzovski:2022vbt}, previously read as a flat ceiling at $\cgg \sim 6\times10^{-2}~\GeV^{-1}$, weakens throughout the chimney once the ALP is allowed to pass the same selection as a $\pi^0$, and disappears for $m_a \approx 120$--$135~\MeV$. The ATLAS monojet search excludes $\cgg$ above $10^{-3}~\GeV^{-1}$ everywhere except in a window around $m_{\pi^0}$ where the produced state decays before escaping, a window a few MeV wide at $3\times10^{-3}~\GeV^{-1}$ and $30~\MeV$ wide at $3\times10^{-2}~\GeV^{-1}$. The ATLAS inclusive photon measurement excludes exactly that window, since the decays it contains are reconstructed as isolated photons with a spectrum harder than the background. Together the two cover the chimney up to $\cgg \simeq 0.15~\GeV^{-1}$, the cutoff of the effective theory for $\Lambda = 1~\TeV$. Above it, the coupling implies colored states below a TeV in the simplest KSVZ completions, which LHC searches exclude. That bound sits at $1.4\times10^{-3}~\GeV^{-1}$ for $N y_Q = 2$ and at most eight times higher for the most generous choice of $N$ and $y_Q$. The pion mass splitting adds an independent, purely kinematic check of the sliver $134.1~\MeV \lesssim m_a \lesssim 135.4~\MeV$ for $10^{-3}~\GeV^{-1} \lesssim \cgg \lesssim 0.1~\GeV^{-1}$. The reach of this observable is set by the $\pm30~\ns$ neutron time-of-flight window imposed in the original analysis; a reanalysis of the raw data could strengthen it considerably. We note that a recent analysis~\cite{Raza:2026nmy} reports additional strong constraints in this region; however, we were unable to reproduce them, and the difference may arise from the treatment of $a$--$\pi^0$ mixing.

All of our coverage of the chimney comes from the LHC, and a low-energy probe such as the KOTO recast proposed in Ref.~\cite{Balkin:2025aer} would confirm it independently. Such a probe matters most at large coupling, where the effective theory is no longer valid at LHC energies and the chimney is closed only by bounds on the UV completion. These are hard to evade, since a large coupling implies a low scale of new physics, but they are model dependent. Kaon decays probe energies far below the cutoff, where the effective description remains valid, and would close this region without reference to the completion.

Beyond the pion, the same mixing occurs wherever the ALP is nearly degenerate with another neutral meson, the $\eta$ and $\eta'$ in particular. Our mass-eigenstate approach extends directly to these cases, and we expect similar chimneys near those masses, where an ALP could still be hiding and where the same boosted signatures at the LHC offer a way to find it.


\begin{acknowledgments}
We thank Misa Toman, Andre Frankenthal, Camilla Mupo, Kiley Kennedy, Patrick Draper, Elliot Maderazo, and Maksym Ovchynnikov for helpful discussions.
We thank Tousif Raza for helpful correspondence regarding Ref.~\cite{Raza:2026nmy}.
This work was supported by the Natural Sciences and Engineering Research Council of Canada (NSERC). 
The work of S.W.L. was supported by the National Science Foundation under Grant No.\ PHY-2544442.
The work of A.M.E. is supported in part by the W.~M.~Keck Foundation through its Bridge Funding Initiative.
\end{acknowledgments}


\appendix 


\section{Derivation of the Pion Mass-Splitting Constraint}
\label{app:tof}

This appendix gives the derivation behind the pion mass-splitting constraint of Sec.~\ref{sec:pi-mass}.

The mass splitting between $\pi^0$ and $\pi^\pm$ is best measured through the speed of the neutron produced in the charge-exchange reaction, $\pi^- p \to n \pi^0$~\cite{Crawford:1990jc}. Low-energy $\pi^-$ stopped in liquid hydrogen bind to protons, forming a bound state of pionic hydrogen, \piH. The \piH{} then decays to $n \pi^0$ through the charge exchange reaction. Assuming the initial \piH{} is at rest, the two-body kinematics fix the neutron speed $\beta_{nr}$ in terms of the masses involved. Measurement of the neutron speed thus determines $m_{\pi^-} - m_{\pi^0}$. Reference~\cite{Crawford:1990jc} measured the neutron time of flight, $t_n = L/(\beta_n c)$, at baselines $L = 3.2$, $7.9$, and $18.1~\m$, and found
\beq
    \label{eq:measured_neutron_speed}
    \beta_{nr}^{(\rm obs)} c = 0.894 266 \pm 0.000063~\mathrm{cm/ns}.
\eeq
The residual motion of the \piH{} atoms, with mean kinetic energy $\overline{T}_{\piH} = 16.2 \pm 1.3~\eV$, spreads the measured speeds $\beta_n$ around $\beta_{nr}$, giving the peak a variance $\sigma_\beta^2 = 2\overline{T}_{\piH}/(3 m_{\piH})$. 

With $a$--$\pi^0$ mixing, \piH{} can decay to either mass eigenstate, $\piH \to n \varphi_i$. Each channel emits the neutron at a different speed $\beta_{nr}^{(i)}$, so the event spectrum becomes a mixture,
\beq
    \frac{\dd N^{\rm mix}}{\dd \beta} = f \frac{\dd N}{\dd \beta}\left(\beta; \beta_{nr}^{(1)}\right) + (1-f) \frac{\dd N}{\dd \beta}\left(\beta; \beta_{nr}^{(2)}\right),
\eeq
where $\dd N/\dd\beta(\beta;\beta_{nr})$ is a speed distribution corresponding to a single eigenstate centered at $\beta_{nr}$, and $f$ is the fraction of decays into $\varphi_1$. 
For sufficiently separated peaks the mixture consists of two distinct peaks; near the $m_a \approx m_\pi$ regime, however, the contributions overlap and the effect is to broaden a single peak, increasing its variance by $\Delta \sigma_\beta^2 = f(1-f)(\Delta\beta)^2$, where $\Delta\beta = \beta_{nr}^{(1)} - \beta_{nr}^{(2)}$.

How far apart the peaks are is set by the mass splitting $\Delta m_\varphi = m_{\varphi_2} - m_{\varphi_1}$: the arrival times of neutrons from the two types of decays differ by
\beq
    \label{eq:peak_separation}
    \Delta\tau = \frac{L\, \Delta \beta}{\big(\beta_{nr}^{(\rm obs)}\big)^2 c} \simeq 17~\mathrm{ns}\,\left(\frac{L}{1~\m}\right)\left(\frac{\Delta m_\varphi}{\MeV}\right).
\eeq
The published spectra of Ref.~\cite{Crawford:1990jc} cover only $\pm\tau_{\rm win}$ around the neutron peak, with $\tau_{\rm win} \simeq 30~\mathrm{ns}$. A second peak outside this window is simply not recorded, and since the overall normalization is not predicted, losing a fraction of events cannot be distinguished from no mixing. The measurement is therefore only sensitive when both peaks lie in the window,
\beq
    \label{eq:window}
    \Delta m_\varphi \lesssim 1.8~\MeV \left(\frac{\tau_{\rm win}}{30~\mathrm{ns}}\right) \left(\frac{1~\m}{L}\right),
\eeq
\ie $|m_a - m_{\pi^0}| \lesssim 0.6~\MeV$ at the shortest baseline and $\lesssim 0.1~\MeV$ at the longest. One can also easily estimate the ceiling of this constraint. Equation~\eqref{eq:mass-splitting} gives the minimum splitting of the two eigenstates at a given coupling. For $\cgg \gtrsim 0.1~\GeV^{-1}$ even this minimum separates the two peaks by more than $30~\mathrm{ns}$, so one of them falls outside the window and the measurement loses sensitivity at any $m_a$.

To quantify the sensitivity we use the published spectra of Ref.~\cite{Crawford:1990jc} directly. The event counts are not published, so for each baseline we digitize the fitted curve over the $\pm\tau_{\rm win}$ window and use it both as the single-peak template and as the spectrum to be fit. The absolute time calibration is not contained in the published curves, so a common shift of the two peaks is profiled, while the template width is held fixed. We compare the two spectra with a binned likelihood, and sum $-2\Delta\ln\mathcal{L}$ over the three baselines. The region excluded at 95\% CL, shown in Fig.~\ref{fig:ksvz-bound}, is $134.1~\MeV \lesssim m_a \lesssim 135.4~\MeV$ and $10^{-3}~\GeV^{-1} \lesssim \cgg \lesssim 0.1~\GeV^{-1}$.


\section{Robustness of UV Constraints}
\label{sec:UV-robust}

In this appendix we consider how robust the bound of Sec.~\ref{sec:UV} is with respect to model variations. 

The direct connection between $m_Q$ and $f$ resulted from a minimal assumption about the KSVZ sector. We can ask whether it is possible to engineer a situation where all the new quarks are decoupled, \ie $m_Q \gg f$. Several attempts with multiple copies of $Q_{L,R}$ and different charge assignments always featured at least one quark with mass scaling as Eq.~\eqref{eq:mass_in_terms_of_cgg}. We argue here that this is generic: one always has a PQ-charged quark at or below $\sim f$, up to the factors of $N$ and $y_Q$ discussed in Sec.~\ref{sec:UV}. An arbitrary renormalizable KSVZ-type model has a mass and Yukawa sector of the form 
\beq
  \mathcal{L}\supset Q_{Li}\left(y_{ij} S + \bar{y}_{ij} S^* + M_{ij}\right) Q_{Rj}^{c} \equiv Q_{Li} \mathcal{M}_{ij} Q_{Rj}^c,
\eeq
where the non-zero terms are dictated by the PQ charge assignments; here $y$ and $\bar{y}$ are Yukawa terms, and $M_{ij}$ are vector-like mass terms. One might hope that one can choose the charges and $M$ to pull up all of the eigenstates such that $m_{Q}\gg f$. This does not work, for the following reason. We want to generate $a G\widetilde{G}$ from this construction, which is achieved by an appropriate chiral rotation of the quarks, one that mimics the PQ rotation for the charge assignment chosen. Such a rotation shifts the QCD theta term by 
\beq
  \Delta \mathcal{L} = \frac{\alpha_s}{8\pi} \left(\arg \det \mathcal{M}\right) G\widetilde{G}.
\eeq
Comparing this to Eq.~\eqref{eq:axion_shift_to_ggdual}, we see that we must have 
\beq
  \frac{1}{2}\arg \det \mathcal{M} = N \frac{a}{v_S}.
\eeq
Since $\det \mathcal{M}$ is a product of mass eigenvalues, this requires that $\det \mathcal{M} \propto e^{i 2N a/v_S}$. Since the only way in which such factors appear is through powers of $\langle S\rangle  = v_S e^{i a/v_S}/\sqrt{2}$, we have $\det \mathcal{M} \propto v_S^n$, where $n\geq 1$ is an integer. That is, $\mathcal{M}$ must have at least one eigenvalue at the scale $v_S \sim f$. 


\bibliographystyle{apsrev4-2}
\bibliography{refs}

\end{document}